# Bloch surface waves at the telecommunication wavelength with Lithium Niobate as top layer for integrated optics.

Tatiana Kovalevich[1*], Djaffar Belharet[1], Laurent Robert[1], Gwenn Ulliac[1], Myun-Sik Kim[2], Hans Peter Herzig[2], Thierry Grosjean[1] and Maria-Pilar Bernal[1]

[1]Département d'Optique P.M. Duffieux, Institut FEMTO-ST, UMR 6174 CNRS, Université Bourgogne Franche-Comté, 15B Avenue des Montboucons, 25030 Besançon Cedex, France; djaffar.belharet@femto-st.fr; laurent.robert@femto-st.fr; gwenn.ulliac@femto-st.fr; thierry.grosjean@univ-fcomte.fr; maria-pilar.bernal@univ-fcomte.fr

[2]Optics & Photonics Technology Laboratory, Ecole Polytechnique Fédérale de Lausanne (EPFL), Rue de la Maladière 71b, Neuchâtel, CH-2000, Switzerland; myun-sik.kim@epfl.ch; hanspeter.herzig@epfl.ch

* Correspondence: tnkovalevich@gmail.com; Tel.: +33-77-125-9867

**Abstract:** Lithium niobate (LN) based devices are widely used in integrated and nonlinear optics. This material is robust and resistive to high temperatures, which makes the LN-based devices table, but challenging to fabricate. In this work we report on the design, manufacturing and characterization of engineered dielectric media with thin film lithium niobate (TFLN) on top for the coupling and propagation of electromagnetic surface waves at the telecommunication wavelengths. The designed one-dimentional photonic crystal (1DPhC) sustains Bloch surface waves at the multilayer/air interface at 1550 nm wavelength with a propagation detected over a distance of 3 mm. The working wavelength and improved BSW propagation parameters open the way for exploration of nonlinear properties of BSW based devices. It is also expected that these novel devices will modify BSW propagation and coupling by external thermal/electrical stimuli due to the improved quality of the TFLN top layer of 1DPhC.

## 1. Introduction

Bloch surface waves (BSWs) have been studied extensively and proved to be efficient in biosensing[1,2], optical trapping[3], vapor sensing[4] and fluorescent enhancement[5]. BSWs have been recently used to study third harmonic generation in thin film of GaAs[6], thus exploiting nonlinear applications. Moreover, the propagation length [7] and dynamics[8] of surface plasmons are also improved using BSWs.

Due to long propagation length and strong field confinement BSWs find applications in the field of integrated optics, thanks to low loss dielectric materials[9]. The fundamental dielectric 2D optical elements, such as 2D-lenses and sub-wavelength focusers[10–12], ridge waveguides[13], 2D prisms and gratings[14] have been studied to manipulate the propagation of BSWs. The results have proven BSWs as a novel candidate for 2D optics. Further, R. Dubey et. al.[15] have studied thoroughly more complex ultra-thin optical components which are the key elements of integrated optics applications. These compounds include 2D disk resonators[16–19], BSWs reflectors[20,21], 2D non-diffracting beam[22] and graphene based in-plane photodetectors[23]. The study contributes to the significant advancement of BSWs based 2D optical integrated systems[24–29].

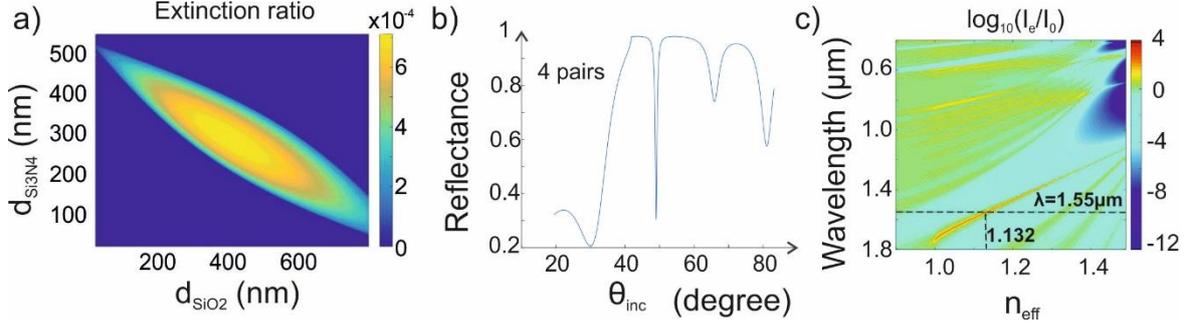

**Figure 1.** (a) Extinction of the band gap as a function of layer thicknesses in 1DPhC, (b) Calculated reflectance at 1550 nm wavelength, (c) Band gap diagram of the 1DPhC.

With the aim of finding new optical functionalities various materials have been used for multilayers and the top (device) layer. For example $Ta_2O_5/SiO_2$[2,30], $SiO_2/Si_3N_4$[10], $TiO_2$[9], graphene[23,31], ZnS[32], $MoS_2$[33] and $LiNbO_3$[34].

Lithium niobate ($LiNbO_3$) is a high refractive index birefringent nonlinear material which has been widely used in integrated optics. Lithium niobate based sensors[35,36], modulators[37], ring and disk resonators[35,38] have been studied, thanks to its ferro-electrical, piezo-electrical, thermo-electrical properties and transparency over a wide wavelength range (350 nm - 5200 nm). Being a top layer in many devices, $LiNbO_3$ provides an enhanced field confinement[35], which can be improved even more by the use of thin films[36].

Recently this material was introduced as a top layer of BSWs based one-dimensional photonic crystal (1DPhC) structures[39]. It has been demonstrated experimentally that modifiable properties of one layer would bring a tunability to the whole 1DPhC[9,34]. Thus one can expect to induce nonlinear behavior and electro/thermo-modulation of BSWs using TFLN as a top layer of the 1DPhC structure.

In this paper, we investigate the coupling and propagation of BSWs on a thin film $LiNbO_3$ based 1DPhC system in order to develop tunable 2D optical systems at telecommunication wavelengths.

## 2. Modeling, Fabrication and Characterization Method

Our 1DPhC design consists of four pairs of $SiO_2$ and $Si_3N_4$ with 450 nm thick X-cut TFLN on the top. The design is optimized in order to work at the telecommunication wavelength of 1550 nm for the applications in integrated optics.

In order to achieve the maximum coupling of surface waves, the thicknesses of $Si_3N_4$ and $SiO_2$ are optimized according to the maximum field extinction ratio (E.R.) for one pair. The details of the maximum field extinction ratio method can be found in ref. [40]. The optimized thicknesses are 250 nm and 450 nm for $Si_3N_4$ and $SiO_2$ respectively. The graphical solution for the E.R. at wavelength 1550 nm, with refractive indices of $n_{Si_3N_4}$ = 1.79 + 0.001i and $n_{SiO_2}$ = 1.44 + 0.001i, is shown in the Fig. (1a).

Pockels effect is a commonly used phenomena for the refractive index change in crystals that lack inversion symmetry, such as $LiNbO_3$[41]. For our optimized multilayer design we chose the configuration which can potentially utilize the largest electro-optic coefficient, and hence leads to a higher change of the refractive index. For $LiNbO_3$ it corresponds to the coefficient $r_{33}$ = 30.8 pm/V at the wavelength of 1550 nm. In this case, for X-cut TFLN, the electrodes should be placed along Y axis of the $LiNbO_3$. Therefore, BSWs should propagate along the Y direction. For the multilayer design modeling we take the ordinary refractive index of $LiNbO_3$ $n_o$= 2.199 + 0.001i.

We exploit Kretschmann configuration to couple BSWs. The details of the coupling set up are explained at the end of this section. The coupling of surface waves depends on the total extinction in the layers and can be verified by observing the reflection dip[33]. For TE polarized incident light, the reflectance dip can be observed at 49_. for the following 1DPhC configuration: BK7 glass /(Si3N4-250 nm/SiO2-450 nm)*4/TFLN-450 nm/air, see Fig. (1b).

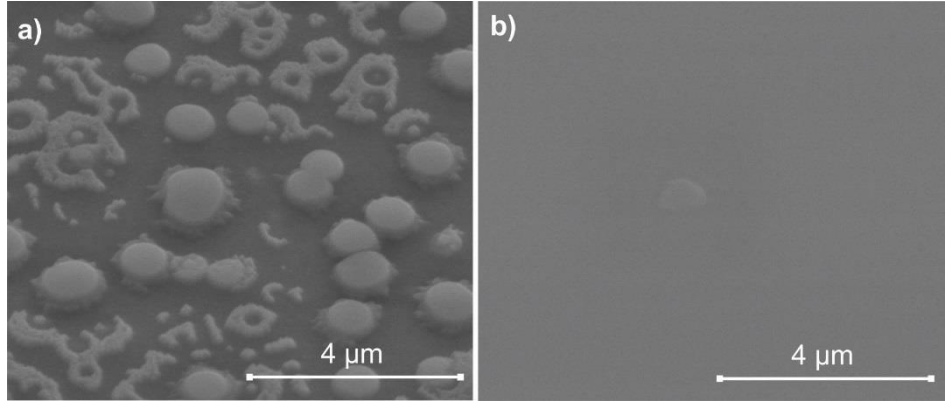

**Figure 2.** SEM images of 1DPhC with TFLN top surface before (a) and after (b) implementation of additional cleaning and temperature stabilization protocols.

The refractive indices used for the calculations are 1.5007, 1.790+0.001i, 1.44+0.001i and 2.199+0.001i for BK7 glass, $Si_3N_4$, $SiO_2$ and $LiNbO_3$, respectively at 1550 nm. The data for refractive indices of all the multilayer materials but lithium niobate (LN) were taken from the refractive index database.

For the TFLN , the refractive index $n_{LiNbO_3}$ is obtained from ellipsometry measurements. For multilayer materials, the imaginary part ($n_i$ = 0.00i) of refractive indices is introduced to consider the losses at the layer interfaces and to locate the reflection dip. For the presented multilayer design, the dispersion curve for 1DPhC is shown in Fig. (1c). The dispersion curve is calculated for the four pairs of alternating $Si_3N_4$ (250 nm thick) and $SiO_2$ (450 nm thick) layers deposited on a BK7 glass and TFLN (450 nm thick) on the top. The external medium is air. We take into consideration the non-homogeneity of the TFLN layer on three-inch wafer. The thickness of TFLN layer varies within 50 nm (from 440 nm to 490 nm) at different positions on the wafer. Considering the thickness variation of the TFLN layer, BSWs can still be excited on 1DPhC stack with aforementioned thicknesses, but at a slightly different incident angle. The dispersion curve of BSWs, in Fig. (1c), would still remain within the photonic band gap of the 1DPhC with a slight shift in the position. The position of the dispersion curve inside the band gap indicates that BSWs cannot propagate in multilayers because of the photonics band gap and total internal reflection on the other side restricts the BSW propagation in an external medium (air in our case). Therefore, light tightly remains confined to the interface of the multilayer and the external medium.

In our previous work [39], we demonstrated several fabrication approaches in order to release a top TFLN surface for external medium sensing and additional nano-structuring, which is essential for integrated optics. Several membrane-based and on-glass-support 1DPhC configurations were proposed in order to excite BSWs at the TFLN/air interface at a wavelength of 473 nm. The samples fabricated on the base of a suspended membrane are fragile and not very comfortable to operate in Kretschmann configuration.

In this study, we focus on an on-glass-support 1DPhC configuration, where a larger area of 1DPhC with TFLN layer is available for BSWs manipulation. The robustness of this configuration facilitates the difficult experimental conditions. By following the fabrication steps described in [39] one indeed can obtain a 1DPhC which sustains a BSW in the visible part of the spectrum. However, these 1DPhCs do not support surface waves at higher wavelengths, particularly at the telecommunication wavelength 1550 nm.

After SEM investigation of the top surface of fabricated 1DPhC as discussed in [39], some residuals of SiO2 were detected (see Fig. 2a). The size of the surface defects is such that the propagation of the BSWs on TFLN surface is possible in the visible part of spectrum, however surface imperfections do not allow surface wave propagation at the desirable wavelength of 1550 nm. These surface imperfections are removed by additional cleaning at the fabrication steps. The SEM image of TFLN surface of 1DPhC fabricated following the new protocols is shown in Fig. 2b. The fabrication process is discussed in the following paragraphs.

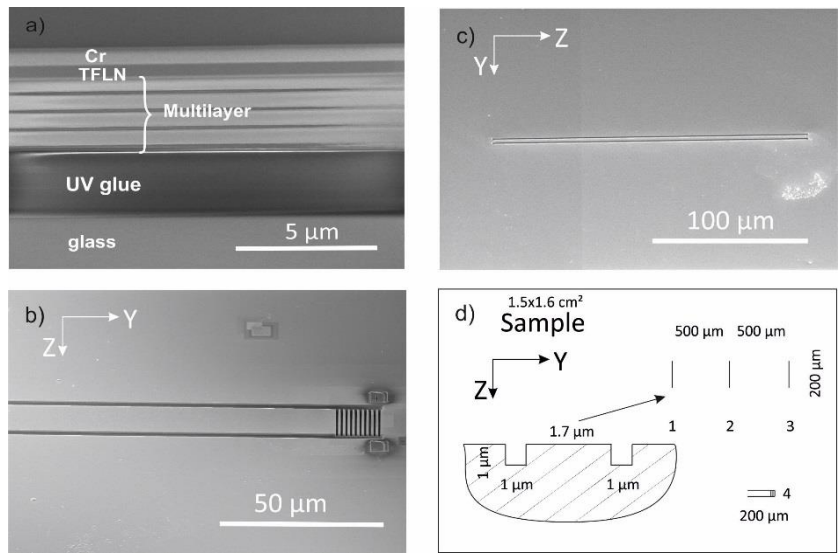

**Figure 3.** (a) FIB-SEM image of the sample cross-section, (b) FIB-SEM image of the waveguide milled along Y axis of LiNbO3, (c) FIB-SEM image of the scatterer, (d) schematics of the scatterers on the sample.

The surface wave existence imposes the requirement of thin layers on the top of 1DPhC. In case of lithium niobate, the deposition of thin layers is challenging. This is because lithium niobate needs to maintain the crystallinity in order to be utilized as a nonlinear material. Nowadays technologies such as sputtering, evaporation or epitaxial growth allow only amorphous or polycrystalline film deposition of LiNbO3 [42]. For sample fabrication, a single-crystal TFLN (450 nm) bonded to a SiO2 layer on Si substrate [43] is used, which avoids light scattering.

The TFLN is prepared using a He+ ion implantation technique[43]. The prepared sample is cleaned with acetone, ethanol and piranha solution. The multilayer stack of alternating SiO2 and Si3N4 are deposited on TFLN by PECVD. The whole structure is bonded to the glass holder (500 $\mu$m) by UV glue and cleaned once again in acetone, ethanol and piranha. The refractive index of UV glue (VITRALIT 6127) is close to the refractive index of the glass holder. The whole stack is protected by S1813 photoresist, and 500 $\mu$m of Si is dry etched by deep reactive ion etching (DRIE). For this etching, the Bosch process[44,45] with the end point detection is employed with SiO2 as a stop layer. The end point detection system based on OES (Optical emission spectroscopy) was used in order to achieve precise etching of silicon when reaching the SiO2 stop layer. Afterwards, the 2 $\mu$m thickness layer of SiO2 is etched by wet etching in a 40% HF solution.

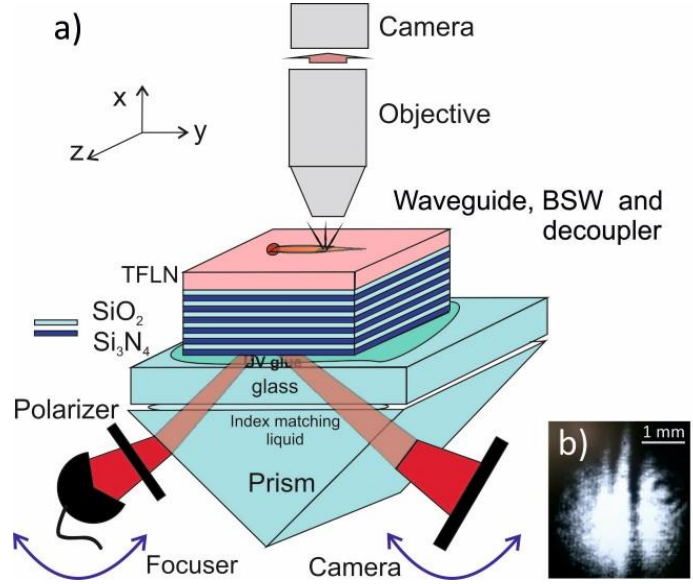

**Figure 4.** (a) Experimental setup, (b) bottom camera image.

As shown in Fig. 2b, a drastic surface improvement is achieved after implementing the above mentioned fabrication steps. This allows the structuring of the top layer of LN to pattern micro optical elements into it. The cross-section of the fabricated sample is shown in Fig. 3a. In order to obtain the SEM image a small opening on the top surface is made by FIB milling. In order to avoid charging effect, an additional layer of 200 nm of Cr is deposited by sputtering. After FIB milling and sample characterization, Cr is removed by a Cr-etch solution.

Figure 3b shows a part of the 200 µm long and 12 µm wide BSW ridge waveguide. The waveguide is milled by FIB milling along the Y crystalline axis of LiNbO$_3$. Thus, it would be possible to exploit the largest electro-optical coefficient of lithium niobate (e.g. by depositing electrodes along Y axis from the both sides of the waveguide). The waveguide is composed of two grooves of 1 µm depth and 1 µm width, respectively. At the end of the waveguide, gratings are milled to perform as a decoupler at 1550 nm wavelength. The grating consists of ten periodic grooves of 1 µm depth and 0.7 µm width. The period of the grating is 1.4 µm.

Additionally, for the BSW propagation estimation, scatterers are fabricated on the top of the 1DPhC surface along the Z crystalline axis of the TFLN. The scatterers are defined by two grooves of 1 µm width and 1 µm depth at the distance of 1.7 µm from each other. The SEM image of a single scatterer is shown in Fig. 3c. Three scatterers at the distance of 0.5 mm from each other are milled by FIB. The schematics of the scatterers' top view and cross section is shown in Fig. 3d.

For the optical characterization, we use the Kretschmann configuration built on a glass prism, see Fig. 4. The light from a tunable laser (1480-1570nm) is focused on the sample through the prism. The illumination is flexible to adjust the size of the beam spot on the top surface of the multilayer chip, where the variable range of the spot size is from 10 µm to 30 µm. The polarization of the incident beam is controlled by a Glan polarizer. On the other side of the prism, the reflected light is collected by an IR camera (camIR 1550, Applied Scintillation). The presence of absorption lines insures the coupling of the BSWs, as shown in Fig. 4b. The reflection dip is observed at an incident angle of 49$^0$, which is very close to the theoretically calculated one. The BSW excited in the TFLN layer propagates along the waveguide and decouples through the grooves at the end of the waveguide. The out-coupled light is recorded by a microscopy imaging system, which combines a long-working distance objective lens (Nikon CF Plan, NA=0.35) and a digital IR camera (Xenix 135 XEVA-2232).

## 3. Results and Discussion

The guided propagation of the BSW on the TFLN is experimentally demonstrated through a 200 μm long waveguide, see Fig. 5. The BSW is excited about 50 μm away from the waveguide. In Fig.5, the two bright spots represent the scattering at the entrance of the waveguide. BSWs propagate through the waveguide and decouple by the gratings at the end of waveguide. The decoupled light is collected by the objective, which can be referred as a bright spot at the end of waveguide. In order to estimate the BSW propagation on the 1DPhC+TFLN surface additional experiments are performed. The BSWs are launched along the Y axis in such a way that scatterer 1, scatterer 2 and scatterer 3 perform as obstacles (see Fig. 3d). Light partially scatters on each scatterer and continues propagating further. It is evident from Fig. 6 that the BSW propagates until scatterer 2. Therefore, having known the distance between scatterers, about 0.5 mm, we may estimate the BSW propagation up to approximately 3 mm of distance on the surface of 1DPhC with TFLN as a top layer.

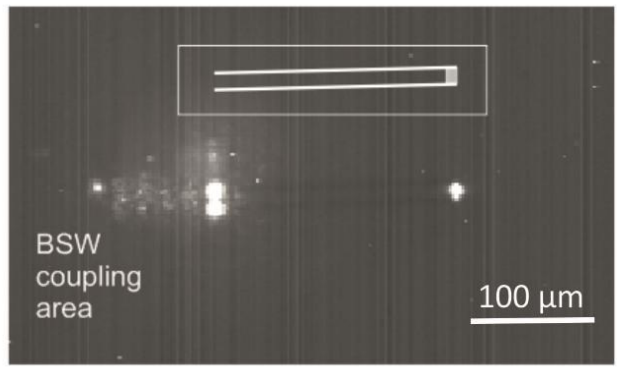

**Figure 5.** The measured far-field image of the out-coupled BSW, which images the BSW coupling area, the entrance of the waveguide, and the grating out-coupler.

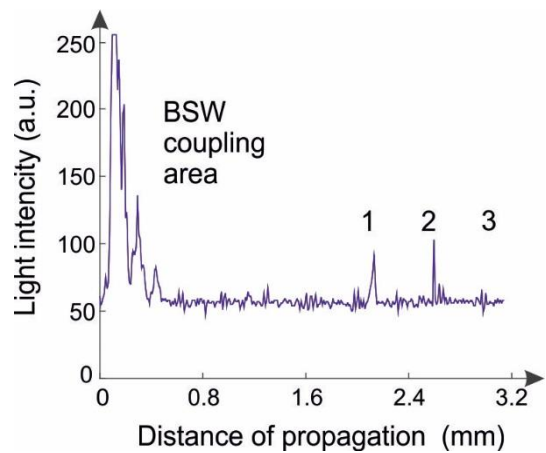

**Figure 6.** Image profile for the BSW excited at 1.5 mm distance from the 1st scatterer.

## 4. Conclusions

In this work, we present the propagation of BSWs along the ridge waveguide patterned into TFLN on the top of 1DPhC. The improved fabrication protocols allow a significant improvement of the surface quality of TFLN deposited on the top of 1DPhC. Therefore, the designed and fabricated multilayer stack (1DPhC+TFLN) couples BSWs at 1550 nm wavelength for integrated and nonlinear optics applications. We demonstrate experimentally the BSWs propagation along Y crystalline axis of LiNbO$_3$ which can be potentially used for electro-optical modulation. Bloch surface wave's propagation of about 3 mm has been measured, for the presented design of 1DPhC with TFLN using a simple microscopy setup, at 1550 nm wavelength.

**Acknowledgments:** This work is funded by the SMYLE program. It has been realized in the context of the Labex ACTION program (contract ANR-11-LABX-01-01). It is supported by the French RENATECH network and its FEMTO-ST technological facility. The authors also thank Raphaël Barbey for multilayer fabrication.

**Author Contributions:** Tatiana Kovalevich performed the experiments, modeled and fabricated samples; Djaffar Belharet contributed to the development of DRIE processes; Laurent Robert contributed to lithography protocols development; Gwenn Ulliac performed FIB milling; Myun-Sik Kim developed and maintained the experimental setup; Hans Peter Herzig, Thierry Grosjean and Maria-Pilar Bernal contributed materials/analysis tools, supervised the work, performed paper corrections.

**Conflicts of Interest:** The authors declare no conflict of interest.

# References

1. Angelini, A.; Enrico, E.; De Leo, N.; Munzert, P.; Boarino, L.; Michelotti F.; Giorgis F.; Descrovi E. Fluorescence diffraction assisted by Bloch surface waves on a onedimensional photonic crystal. New J. Phys. 2013, 15, 073002.
2. 2. Konopsky, V.N.; Alieva, E.V. Photonic crystal surface waves for optical biosensors. Anal. Chem. 2007, 79, 4729-473.
3. 3. Shilkin, D.A.; Lyubin, E.V.; Soboleva, I.V.; Fedyanin, A.A. Direct measurements of forces induced by Bloch surface waves in a one-dimensional photonic crystal. Opt. Lett. 2015, 40, 4883-4886.
4. 4. Michelotti, F.; Sciacca, B.; Dominici, L.; Quaglio,M.; Descrovi, E.; Giorgis, F.; Geobaldo, F. Fast optical vapor sensing by Bloch surface waves on porous silicon membranes. Phys. Chem. Chem. Phys. 2010, 12, 502-506.
5. 5. Michelotti, F.; Sciacca, B.; Dominici, L.; Quaglio, M.; Descrovi, E.; Giorgis, F.; Geobaldo, F. Fluorescence diffraction assisted by Bloch surface waves on a one dimensional photonic crystal. New J. Phys. 2013, 15, 073002.
6. Konopsky, V.N.; Alieva, E.V.; Alyatkin, S.T.; Melnikov, A.A.; Chekalin, S.V. Phase-matched third-harmonic generation via doubly resonant optical surfacemodes in 1D photonic crystals. Light Sci Appl. 2016, 5, 16168.
7. Konopsky, V.N.; Basmanov, D.V.; Alieva, E.V.; Dolgy, D.I.; Olshansky, E.D.; Sekatskii, S.K.; Dietler, G. Registration of long-range surface plasmon resonance by angle-scanning feedback and its implementation for optical hydrogen sensing. New J. Phys. 2009, 11, 063049.
8. Zhang, D.; Wang, R.; Xiang, Y.; Kuai, Y.; Kuang, C.; Badugu, R.; . . . Lakowicz, J.R. Silver Nanowires for Reconfigurable Bloch Surface Waves. ACS Nano 2017, 11, 10446–10451.
9. Dubey, R.; Barakat, E.; Häyrinen, M.; Roussey, M.; Honkanen, S.; Kuittinen, M.; Herzig, H.P. Experimental investigation of the propagation properties of Bloch surface waves on dielectric multilayer platform. J. Eur. Opt. Soc. 2017, 13, 1–9.
10. Yu l.; Barakat E.; Sfez T.; Hvozdara L.; Di Francesco J.; Herzig H.P. Manipulating Bloch surface waves in 2D: a platform concept-based flat lens. LIGHT-SCI APPL 2014, 3, 124.
11. Kim, M.-S.; Lahijani, B.V.; Descharmes, N.; Straubel, J.; Negredo, F.; Rockstuhl, C.; Häyrinen, M.; Roussey, M.; Kuittinen, M.; Herzig, H.P. Subwavelength focusing of Bloch surface waves. ACS Photonics 2017, 4, 1477–1483.
12. Kim, M.-S.; Barakat, E.; Dubey, R.; Scharf, T.; Herzig, H.P. Nondiffracting Bloch surface wave: 2D quasi-Bessel-Gauss beam. In Proceeding of the OSA CLEO Pacific Rim Conference, Busan, Korea, 24–28 August 2015.
13. Descrovi, E.; Sfez, T.; Quaglio, M.; Brunazzo, D.; Dominici, L.; Michelotti, F.; Herzig, H.P.; Martin, O,; Giorgis, F. Guided Bloch surface waves on ultra-thin polymeric ridges. Nano Lett. 2010, 10, 2087-2091.
14. Yu, L. Near-field Imaging: Investigations on Bloch Surface Wave Based 2D Optics and the Development of Polarization-retrieved characterization. Doctoral dissertation, EPFL 2013.

15. Dubey, R.; Barakat, E.; Herzig, H.P. Bloch Surface Waves Based Platform for Integrated Optics. In Proceeding of the IEEE Photonics Conference, 1092–8081, Reston, VA, USA, 4–8 October 2015.
16. Dubey, R.; Lahijani, B.V.; Barakat, E.; Häyrinen, M.; Roussey, M.; Kuittinen, M.; Herzig, H.P. Near-field characterization of a Bloch-surfacewave-based 2D disk Resonator. Opt. Lett. 2016, 41, 2087-2091.
17. Lahijani, B.V.; Ghavifekr, H.B.; Dubey, R.; Kim, M.-S.; Vartiainen, I.; Roussey, M.; Herzig, H.P. Experimental demonstration of critical coupling of whispering gallery mode cavities on Bloch surface wave platform. Opt. Lett. 2017, 42, 5137–5140.
18. Dubey, R.; Lahijani, B.V.; Kim, M.-S.; Barakat, E.; Häyrinen, M.; Roussey, M.; Kuittinen, M.; Herzig, H.P. Near-field investigation of Bloch surface wave based 2D optical components. In Proceeding of the SPIE Photonics West 10106, San Francisco, CA, USA, 16 February 2017.
19. Dubey, R. Near-Field Characterization of Bloch Surface Waves Based 2D Optical Components. Ph.D. Thesis, École Polytechnique Fédérale de Lausanne, Lausanne, Switzerland, 2017.
20. Dubey, R.; Lahijani, B.V.; Häyrinen, M.; Roussey, M.; Kuittinen, M.; Herzig, H.P. Ultra-thin Bloch surface waves based reflector at telecommunication wavelength. Photonics Res. 2017, 5, 494–499.
21. Dubey, R.; Lahijani, B.V.; Roussey, M.; Herzig, H.P. Waveguide grating as a Bragg mirror on Bloch surface waves based platform for 2D integrated optics applications. In Proceeding of SPIE Optics + Photonics 10730-22, San Diego, CA, United States, September 2018.
22. Kim, M.-S.; Dubey, R.; Barakat, E.; Herzig, H.P. Nano-thin 2D axicon generating nondiffracting surface waves. In Proceedings of the Optical MEMS and Nanophotonics (OMN), Singapore, 31 July–4 August 2016.
23. Dubey, R.; Marchena, M.; Vosoughi Lahijani, B.; Kim, M.-S.; Pruneri, V.; Herzig, H.P. Bloch Surface Waves Using Graphene Layers: An Approach toward In-Plane Photodetectors. Appl. Sci. 2018, 8, 390.
24. Herzig, H. P.; Barakat, E.; Dubey, R.; Kim, M.-S. Optics in 2D Bloch surface wave phenomena and applications. 15th Workshop on Information Optics (WIO), 1-3, Barcelona, Spain, 2016. [10.1109/WIO.2016.7745573]
25. Dubey, R.; Barakat, E.; Kim, M.-S.; Herzig, H. P. Near-field characterization of 2D disk resonator on Bloch surface wave platform. 14th International Conference on Near-field Optics, Nanophotonics, and Related Techniques, Hamamatsu, Japan, 2016. [infoscience.epfl.ch/record/221501]
26. Herzig, H. P.; Barakat, E.; Yu, L.; Dubey, R. Bloch surface waves, a 2D platform for planar optical integration. 13th Workshop on Information Optics (WIO), Neuchatel, Switzerland, 2014. [10.1109/WIO.2014.6933280]
27. Dubey, R.; Barakat, E.; Herzig, H. P. Near Field Investigation of Bloch Surface Based Platform for 2D Integrated Optics. PIERS Progress In Electromagnetics Research Symposium, Prague, Czech Republic, 2015. [infoscience.epfl.ch/record/210317]
28. Kim, M.-S.; Dubey, R.; Barakat, E.; Herzig, H. P. Exotic optical elements generating 2D surface waves. EOS Topical Meeting on Trends in Resonant Nanophotonics, Berlin, Germany, 2016. [infoscience.epfl.ch/record/221568]
29. Dubey, R.; Barakat, E.; Herzig, H. P. Bloch Surface Based Platform for Optical Integration. TOM 5 – Metamaterials, Photonic Crystals and Plasmonics: Fundamentals and Applications, Berlin, Germany, 2014. [infoscience.epfl.ch/record/207822]
30. Konopsky, V.N.; Karakouz, T.; Alieva, E.V.; Vicario, C.; Sekatskii, S.K.; Dietler, D. Photonic crystal biosensor based on optical surface waves. Sensors 2013, 13, 2566-2578.
31. Xiang, Y.; Guo, J.; Dai, X.; Wen, S.; Tang, D. Engineered surface Bloch waves in graphene-based hyperbolic metamaterials. Opt. Express 2014, 22, 3054-3062.
32. Sekatskii, S.K.; Smirnov, A.; Dietler, G.; Nur E. Alam, M.; Vasiliev, M.; Alameh, K. Photonic crystal-supported long-range surface plasmon-polaritons propagating along high-quality silver nanofilms. Appl. Sci. 2018, 8, 248.
33. Lerario, G.; Ballarini, D.; Dominici, L.; Fieramosca, A.; Cannavale, A.; Holwill, M.; Kozikov, A.; Novoselov, K.S.; Gigli, G. Bloch Surface Waves for MoS2 Emission Coupling and Polariton Systems. Appl. Sci. 2017, 7, 1217.

34. Kovalevich, T.; Ndao, A.; Suarez, M.; Häyrinen, M.; Roussey, M.; Kuittinen, M.; Grosjean, T.; Bernal, M.P. Tunable Bloch surface waves in anisotropic photonic crystals based on lithium niobate thin films. Opt. Lett. 2016, 41, 5616-5619.
35. Chen, L.; Reano, R.M. Compact electric field sensors based on indirect bonding of lithium niobate to silicon microrings. Opt. Express 2012, 20, 4032-4038.
36. Qiu,W.; Lu, H.; Baida, F.I.; Bernal,M.P. Ultra-compact on-chip slot Bragg grating structure for small electric field detection. Photon. Res. 2017, 5, 212-218.
37. Wang, C.; Zhang, M.; Stern, B; Lipson, M.; Lončar, M. Nanophotonic lithium niobate electro-optic modulators. Opt. Express 2018, 26, 1547-1555.
38. Wang, J.; Bo, F.;Wan, S.; Li,W.; Gao, F.; Li, J.; Zhang, G.; Xu, J. High-Q lithium niobate microdisk resonators on a chip for efficient electro-optic modulation. Opt. Express 2015, 23, 23072-23078.
39. Kovalevich, T.; Kim, M.S.; Belharet, D.; Robert, L.; Herzig, H.P.; Grosjean, T.; Bernal, M.P. Experimental evidence of Bloch surface waves on photonic crystals with thin film LiNbO3 as a top layer. Photon. Res. 2017, 5, 649-653.
40. Konopsky, V.N. Plasmon-polariton waves in nanofilms on one-dimensional photonic crystal surfaces. Photon. Res. 2010, 12, 093006.
41. Kunz, K.; Luebbers, R.J. The finite difference time domain method for electromagnetics. CRC press, 1993.
42. Bezpaly, A.D.; Shandarov, V.M. Optical formation of waveguide elements in photorefractive surface layer of a lithium niobate sample. Phys. Procedia. 2017, 86, 166 - 169.
43. Han, H.; Cai, L.; Hu, H. Optical and structural properties of single-crystal lithium niobate thin film. Opt. Mater. 2015, 42, 47-51.
44. Lärmer, F.; Schlip, A, A Method of Anisotropically Etching Silicon, Licensed from Robert Bosch GmbH: US Patent No. 5,501,893 1996.
45. Schaepkens, M.; Bosch, R.C.M.; Standaert, T.E.F.M.; Oehrlein, G.S.; Cook, J.M. Influence of reactor wall conditions on etch processes in inductively coupled fluorocarbon plasmas J. Vac. Sci. Technol. A 1998, 16, 2099.